\documentclass[aps,prd,twocolumn,superscriptaddress,floatfix,preprintnumbers,nofootinbib,showpacs]{revtex4}

\usepackage{color}
\usepackage[dvips]{graphicx}

\newcommand{\lsim}{\mathrel{\mathop{\kern 0pt \rlap {\raise.2ex\hbox{$<$}}}\lower.9ex\hbox{\kern-.190em $
\sim$}}}

\newcommand{\gsim}{\mathrel{\mathop{\kern 0pt \rlap{\raise.2ex\hbox{$>$}}}\lower.9ex\hbox{\kern-.190em $\sim
$}}}

\newcommand{\be}{\begin{equation}}
\newcommand{\ee}{\end{equation}}
\newcommand{\beqarr}{\begin{eqnarray}}
\newcommand{\eeqarr}{\end{eqnarray}}

\begin{document}
\preprint{DFTT 18/2007}
\preprint{KIAS--P07045}

\title{Zooming in on light relic neutralinos by direct detection and measurements of galactic antimatter}


\author{A. Bottino}
\email{bottino@to.infn.it}
\affiliation{Dipartimento di Fisica Teorica, Universit\`a di Torino \\
Istituto Nazionale di Fisica Nucleare, Sezione di Torino \\
via P. Giuria 1, I--10125 Torino, Italy}
\author{F. Donato}
\email{donato@to.infn.it}
\affiliation{Dipartimento di Fisica Teorica, Universit\`a di Torino \\
Istituto Nazionale di Fisica Nucleare, Sezione di Torino \\
via P. Giuria 1, I--10125 Torino, Italy}
\author{N. Fornengo}
\email{fornengo@to.infn.it}
\affiliation{Dipartimento di Fisica Teorica, Universit\`a di Torino \\
Istituto Nazionale di Fisica Nucleare, Sezione di Torino \\
via P. Giuria 1, I--10125 Torino, Italy}
\author{S. Scopel}
\email{scopel@kias.re.kr}
\affiliation{Korea Institute for Advanced Study \\
Seoul 130-722, Korea}
\date{\today}

\begin{abstract}

The DAMA Collaboration has recently analyzed its
data of the extensive  WIMP direct search (DAMA/NaI) which detected an annual modulation,
by taking into account the channelling
effect which occurs when an ion  traverses a detector with a crystalline structure. Among
possible implications, this Collaboration has considered the case of a coherent
WIMP-nucleus interaction and then derived  the form of the
annual modulation region in the plane of the WIMP-nucleon cross section versus the WIMP mass,
 using a specific modelling for the channelling effect.
In the present paper we first show that light neutralinos
fit the annual modulation region also when channelling  is taken into account.
To discuss the connection with indirect signals consisting in galactic antimatter, in our
analysis we pick up a
specific galactic model, the cored isothermal-sphere. In this scheme
we determine the sets of supersymmetric
models selected by the annual modulation regions and  then  prove that these sets are
compatible with the available data on galactic antiprotons. We comment on implications
when other galactic distribution functions are employed. Finally, we show that future
measurements on galactic antiprotons and antideuterons will be able to shed further light on
the populations of light neutralinos singled out by the annual modulation data.
\end{abstract}


\maketitle

\section{Introduction}
\label{sec:intro}

\begin{figure}[t] \centering
\vspace{-20pt}
\includegraphics[width=1.0\columnwidth]{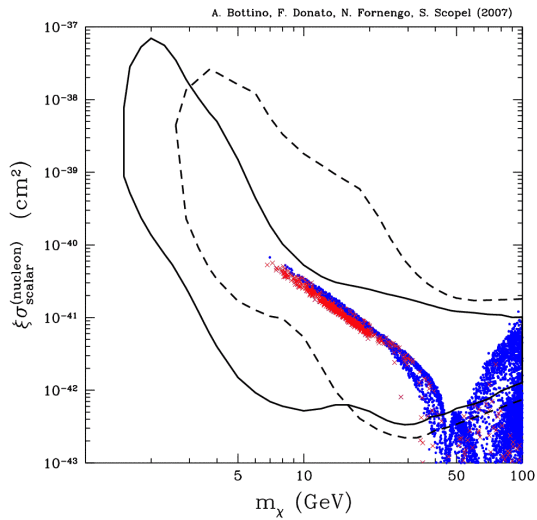}
\vspace{-30pt}
  \caption{WIMP--nucleon scattering cross-section as a function of the WIMP mass.
  The solid (dashed) line denotes the annual modulation region derived by the DAMA Collaboration with
  (without) the inclusion of the channeling effect. The two regions contain points where the likelihood-
  function values differ more than 4$\sigma$ from the null hypothesis (absence of modulation).
  These regions are obtained by varying the WIMP
  galactic distribution function (DF) over the set considered in Ref. \cite{bcfs}
  and by taking into account
  other uncertainties of different origins \cite{damalast}. The scatter plot
  represents supersymmetric configurations calculated with the supersymmetric model
  summarized in the Appendix. The (red) crosses denote configurations
  with a neutralino relic abundance which matches the WMAP cold
  dark matter amount ($0.092 \leq \Omega_{\chi} h^2 \leq 0.124$), while the
  (blue) dots refer to configurations where the neutralino is subdominant
  ($\Omega_{\chi} h^2 < 0.092$).}
\label{fig:00}
\end{figure}

In a recent paper  \cite{damalast} the DAMA Collaboration has
analyzed the data of its extensive  WIMP direct search (DAMA/NaI)
\cite{dama} which measured an annual modulation effect at 6.3
$\sigma$ C. L., by taking into account the channelling effect.  This
effect occurs when an ion  traverses a detector with a crystalline
structure  \cite{drob}. In Ref. \cite{damalast} implications of
channelling  have been  discussed in terms of a specific modelling
of this effect for the case of the DAMA NaI(Tl) detector; it is
shown that the occurrence of  channelling  makes the response of
this detector to WIMP-nucleus interactions more sensitive than in
the case in which channelling is not included. Therefore, when
applied to a WIMP with a coherent interaction with nuclei, the
inclusion of the channelling effect implies that the annual
modulation region, in the plane of the WIMP-nucleon cross section
versus the WIMP mass, is considerably modified as compared to the
one derived without including channelling. The extent of the
modification depends on the specific model--dependent procedure
employed in the evaluation of the channeling effect \cite{damalast}.

These properties are shown in Fig. \ref{fig:00}, where the quantity
$\sigma^{\rm nucleon}_{\rm scalar}$ denotes the WIMP-nucleon scalar
cross-section, $\xi = \rho_{\rm WIMP}/\rho_0$ is the WIMP local
fractional matter density and $m_{\chi}$ is the WIMP mass. The
dashed line denotes the annual modulation region derived by the DAMA
Collaboration without including the channeling effect \cite{dama}.
The solid line shows the annual modulation region derived by the
same Collaboration when the channeling effect is included as
explained in Ref. \cite{damalast}.

The regions displayed in Fig. \ref{fig:00} are derived by varying
the WIMP galactic distribution function (DF) over the set considered
in Ref.\cite{bcfs} and by taking into account other uncertainties of
different origins \cite{damalast,nota1}. Fig. \ref{fig:00} shows
that   the  effect of taking channelling into account is  that
 the annual modulation region modifies its contour with an extension towards lighter
WIMP masses. Most remarkably, for WIMP masses $\lsim $ 30 GeV, the
WIMP-nucleon cross section involved in the annual modulation effect
decreases sizeably, up to  more than an order of magnitude. As
mentioned before, the specific shape of the annual modulation region
depends on the way in which channelling is modelled \cite{damalast}.

These features are  of great importance for a specific
dark matter candidate, the light neutralino,
which was extensively
investigated in Refs. \cite{lowneu,lowdir,ind}. Actually,
in these papers  we analyzed  light neutralinos, {\it i.e.}
neutralinos with a mass $m_{\chi} \lsim 50$ GeV, which arise naturally
in supersymmetric models where gaugino mass parameters are not
related by a GUT--scale unification condition.
In Refs. \cite{lowneu,lowdir} it is proved that,
when R-parity conservation is assumed,
these neutralinos are of great relevance for the DAMA/NaI annual modulation effect.
In these papers it is also shown that in MSSM without gaugino 
mass unification the lower limit of the neutralino mass is 
$m_{\chi} \gsim$ 7 GeV \cite{hp}.

In Fig. \ref{fig:00}, superimposed to the annual modulation regions is  the scatter plot of
the supersymmetric configurations of our model, whose features are summarized in the Appendix.
One sees that, also when the channeling effect is taken into account,
the light neutralinos of our supersymmetric model  fit quite well the
annual modulation region.

In the present paper we consider the phenomenological consequences
for light neutralinos when the annual modulation region is the one
indicated by the solid line in Fig.\ref{fig:00}. More specifically
we examine the properties of our supersymmetric population of light
relic neutralinos in terms of the possible antimatter components
generated by their pair annihilation in the galactic halo.

To do this, we have to resort to a specific form for the WIMP DF.
We take as our representative DF a standard cored isothermal-sphere
model, though we do not  mean to associate
 to this model prominent physical motivations as compared to other forms of DFs.
Analyses similar to the one we present here for the cored isothermal-sphere
can be developed for  other
galactic models. We will comment about some of them, selected among those
considered in Ref. \cite{bcfs} (we will follow the denominations of this
Reference to classify our DFs).

The scheme of
the present paper is the following. In Sect. II, we show how  the model
presented in Refs.  \cite{lowneu,lowdir,ind} fits
the DAMA/NaI annual modulation regions of Ref. \cite{damalast} for the case of the
cored isothermal-sphere model.
In Sect. III we combine
these results with constraints derivable from available data on
cosmic antiprotons; we also discuss the sensitivity of upcoming
measurements on cosmic antiprotons for investigating
the neutralino populations selected by the annual modulation regions.
Complementary investigations by measurements of galactic antideuterons
are presented in Sect. IV. Conclusions are drawn in Sect.V.
The main features of the
supersymmetric scheme adopted here are summarized in the Appendix.

\section{The annual modulation region in various halo models}

\begin{figure*}[t] \centering
\vspace{-20pt}
\includegraphics[width=2.0\columnwidth]{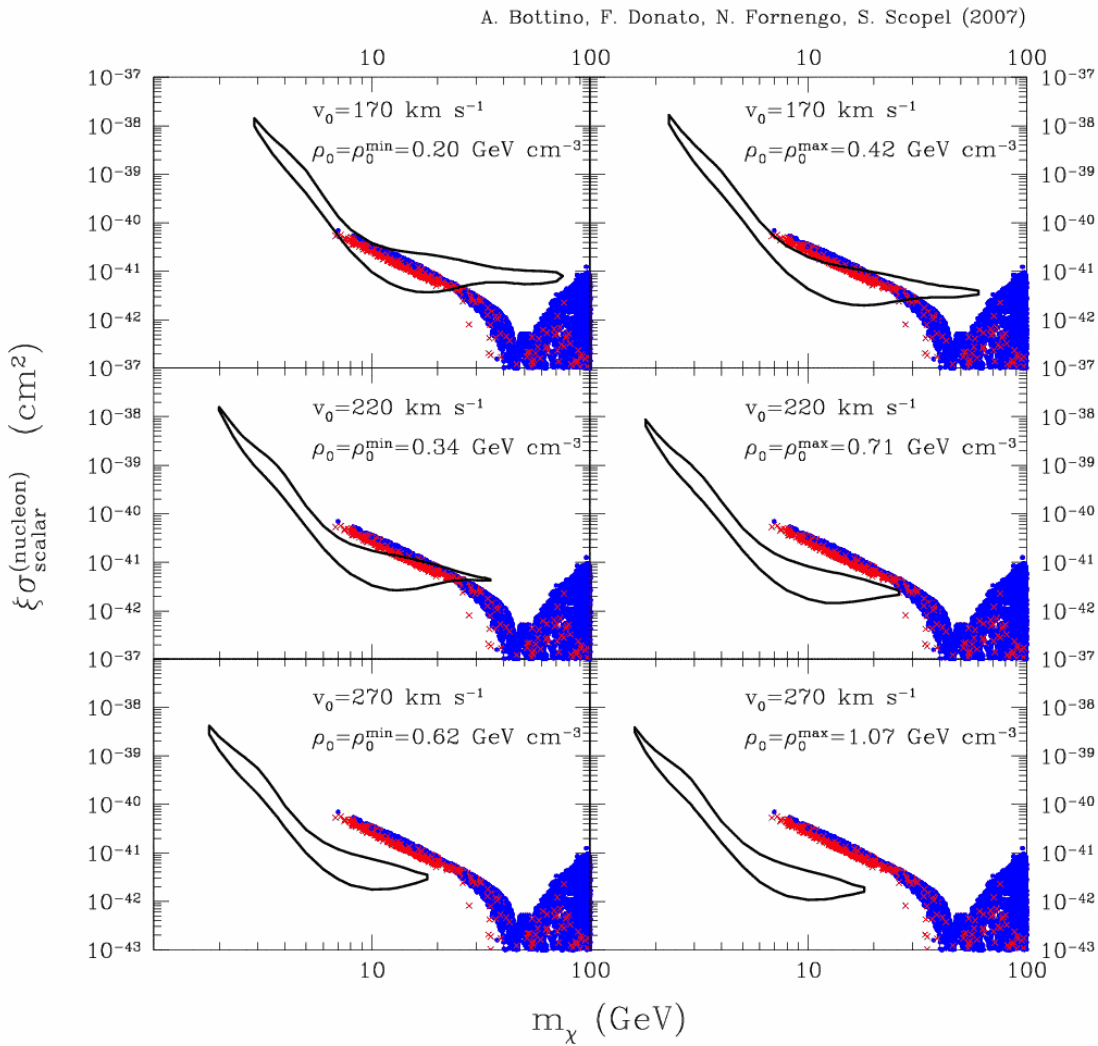}
\vspace{-30pt}
  \caption{WIMP--nucleon scattering cross-section as a function of the WIMP
  mass. The solid contours denote the DAMA/NaI annual modulation regions for a
  cored isothermal halo, derived by including the channeling effect with the model explained in Ref.
\cite{damalast}. The different panels refer to different galactic
halo--model parameters, according to the analysis of Ref.
\cite{bcfs}: $v_{0}$ is the
  local rotational velocity, $\rho_{0}$ is the local dark matter density. The
  scatter plot shows the configurations for neutralino--nucleon scattering in
  gaugino non--universal supersymmetric models. The (red) crosses denote
  configurations with a neutralino relic abundance which matches the WMAP cold
  dark matter amount ($0.092 \leq \Omega_{\chi} h^2 \leq 0.124$), while the
  (blue) dots refer to configurations where the neutralino is subdominant
  ($\Omega_{\chi} h^2 < 0.092$).}
\label{fig:01}
\end{figure*}

As mentioned above, in the present paper we take the cored isothermal sphere
as the representative
model for our detailed evaluations. Similar analyses can be developed for  other
galactic models; we will comment about some of them.  The density profile of the
cored--isothermal sphere (denoted as Evans logarithmic model, or A1 model,  in
Ref.  \cite{bcfs}) is:

\begin{equation}
\rho(r) = \frac{v_0^2}{4 \pi G}\frac{3 R_c^2 + r^2}{(R_c^2 + r^2)^2}\, ,
\label{isot}
\end{equation}

\noindent
where $G$ is the Newton's constant, $v_0$ is the local value of the rotational velocity and
$R_c$ is the core radius.

The value $R_c = 5$ kpc will be used for the core radius.
For the parameter $v_0$ we will consider the values $v_0 = 170, 220, 270$ km sec$^{-1}$,
which represent the minimal, central and maximal values of $v_0$ in its physical range  \cite{koch}.
For each value of $v_0$, we will consider the minimal and the maximal values of the local
dark matter density,  $\rho_0^{\rm min}$ and $\rho_0^{\rm max}$, as determined in Ref. \cite{bcfs}.
Then, specifically, we will discuss the following sets of values: i)  $v_0 = 170$ km sec$^{-1}$ with
$\rho_0^{\rm min} = 0.20$ GeV cm$^{-3}$ and  $\rho_0^{\rm max} = 0.42$ GeV cm$^{-3}$;
ii) $v_0 = 220$ km sec$^{-1}$ with
$\rho_0^{\rm min} = 0.34$ GeV cm$^{-3}$ and  $\rho_0^{\rm max} = 0.71$ GeV cm$^{-3}$;
iii)  $v_0 = 270$ km sec$^{-1}$ with
$\rho_0^{\rm min} = 0.62$ GeV cm$^{-3}$ and  $\rho_0^{\rm max} = 1.07$ GeV cm$^{-3}$.

Now, we turn to a comparison of the DAMA/NaI annual modulation regions of Ref.
\cite{damalast} with the theoretical predictions of our supersymmetric model.
Fig. \ref{fig:01} displays the DAMA annual modulation regions in the case of the A1 model
\cite{nota2};
the various insets correspond to the representative values of the parameters
$v_0$ and $\rho_0$ previously defined.  The  regions of Fig. \ref{fig:01} are derived by the DAMA Collaboration from
their data of Ref.  \cite{dama}, taking into account the channelling effect and
under the hypothesis that the WIMP-nucleus interaction is coherent. They
represent regions where the likelihood-function values differ more than
4$\sigma$ from the null hypothesis (absence of modulation) \cite{info}. The scatter plot
is the same as in Fig. \ref{fig:00}. It is remarkable that light neutralinos are able to provide a good fit to the
experimental data. This occurs for values of $v_0$ and $\rho_0$ which are in the low-medium
side of their own physical ranges, {\it i.e.} $v_0 \simeq$ (170 -- 220) km sec$^{-1}$
and $\rho_0 \simeq (0.2 - 0.4)$ GeV cm$^{-3}$. The light neutralinos involved in this fit stay
in the mass range $m_{\chi} \simeq (7 - 30)$ GeV. We remark that in case the channelling is not included, the DAMA regions 
would be sizeably displaced as compared to the ones displayed in Fig. 2, 
similarly to what is shown in Fig. 1. An example of the effect of including 
channelling in the determination of the annual modulation region, when the 
A1 model for the DF is taken, is explicitly displayed in Fig. 7 of Ref. [1]. As 
already remarked before, in connection with Fig. 1, for WIMP masses $\leq$ 
30 GeV, the WIMP-nucleon cross section involved in the annual modulation effect 
decreases sizeably, when channeling is included, up to more than an order of 
magnitude. This implies that, not including channelling, the fit of the experimental data with light 
neutralinos would require values of $\rho_0$ 
($\rho_0 \simeq (0.6 - 1)$ GeV cm$^{-3}$) higher than the ones previously 
derived. Also $v_0$ would be in the high side of its physical range. These 
properties are of relevance for the implications which will follow.

When the channelling effect is taken into account and no rotation of the halo is
considered, it turns out  \cite{info} that the features of the annual modulation
region in the $m_{\chi} -  \xi \sigma^{\rm nucleon}_{\rm scalar}$ plane do not
differ much  when the galactic DF is varied, for many of the galactic DFs
considered in Ref. \cite{bcfs}. Thus, for instance, for a matter density with a
Navarro-Frenk-White profile (A5 model of Ref. \cite{bcfs}) or for an isothermal
model with a non-isotropic velocity dispersion (B1 model of Ref. \cite{bcfs})
the physical situation is very similar to the one depicted in Fig. \ref{fig:01}.
However, in
the case  of DFs with triaxial spatial distributions (within the class D  of
Ref. \cite{bcfs}) and for models with a corotating halo there can be an
elongation of the annual modulation region towards heavier masses
\cite{damalast}. Further insight into the properties of light neutralinos are expected
from the future results of the DAMA/LIBRA experiment \cite{dama}.

We wish also to stress that the distribution
of WIMPs in the Galaxy could deviate from the models mentioned above, mainly
because of the presence of streams. For modification of the annual modulation
region in these instances see Ref.\cite{dama}.
 It is worth mentioning that in the  numerical derivation given above, also
uncertainties of other origin may intervene. Suffice it to mention that the
sizeable uncertainties which affect strength of the coupling of the neutralino
to the nucleon \cite{noi}.

In this paper, among the searches for WIMP direct detection we discuss
only the DAMA/NaI experiment, since this is the only experiment
having at present the capability to measure 
the annual modulation effect,
which is a  distinctive feature for discriminating the signal against the
background in a WIMP direct search \cite{freese}. For updated reviews about
other experiments of WIMP direct detection, see Ref. \cite{review}.

\section{Galactic antiprotons}

\begin{figure*}[t] \centering
\vspace{-20pt}
\includegraphics[width=2.0\columnwidth]{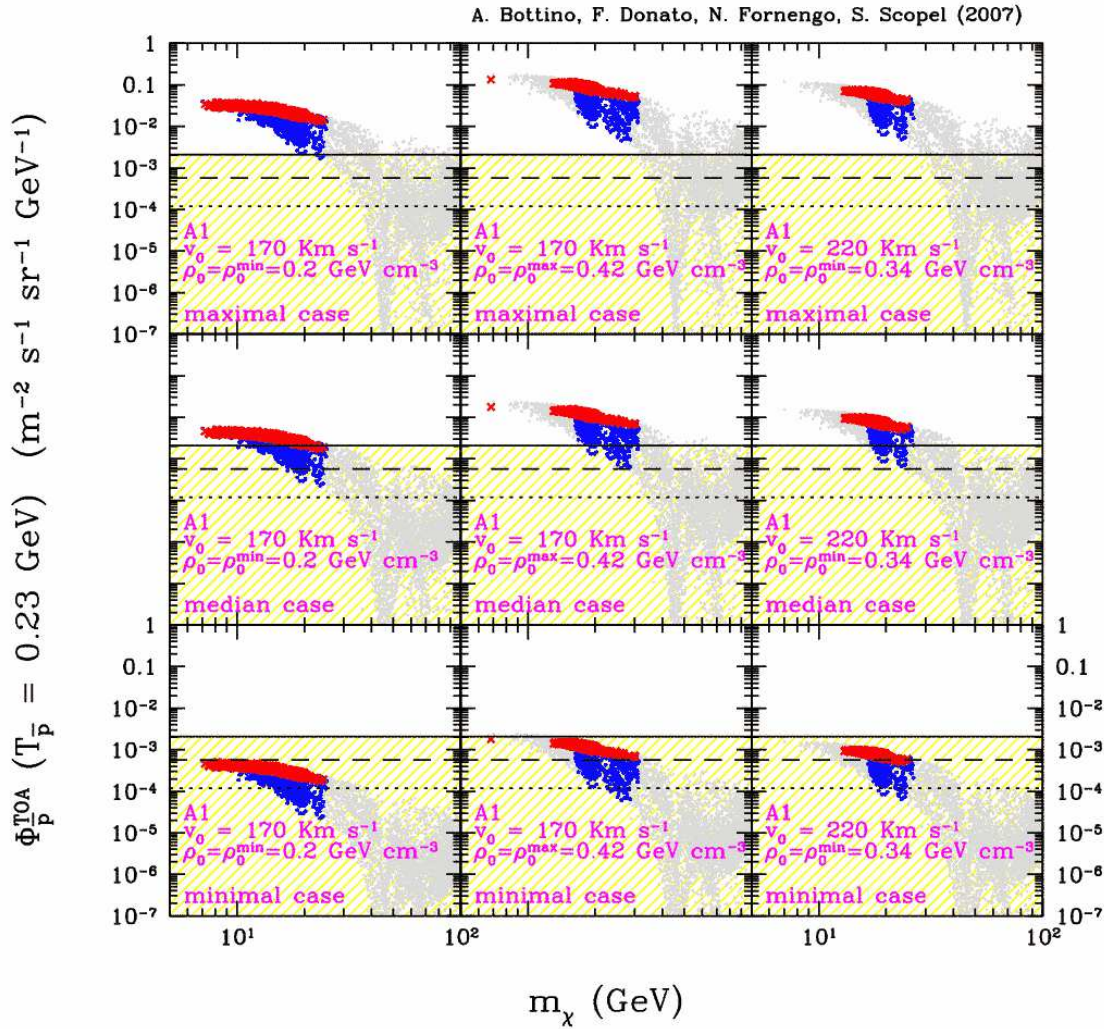}
\vspace{-30pt}
  \caption{Antiproton flux at $\bar p$ kinetic energy $T_{\bar p}=0.23$ GeV, as
  a function of the WIMP mass and for a cored isothermal halo. Each raw
  correspond to a different set of cosmic--rays propagation parameters: the
  upper, median and lower rows refer to the set which provides the maximal,
  median and minimal antiproton flux, according to the analysis of Ref.
  \cite{pbar}. The light gray points denote configurations with a
  neutralino--nucleon scattering cross section outside the corresponding
  DAMA/NaI allowed region. The bold (colored) points refer to configurations
  compatible with the DAMA/NaI regions. These last points are further
  differentiated as follows: (red) crosses denote configurations with a
  neutralino relic abundance which matches the WMAP cold dark matter amount
  ($0.092 \leq \Omega_{\chi} h^2 \leq 0.124$), while (blue) dots refer to
  configurations where the neutralino is subdominant ($\Omega_{\chi} h^2 <
  0.092$). The solid horizontal line shows the maximal allowable amount of
  antiprotons in the BESS data \cite{bess} over the secondary component; the
  dashed and dotted lines denote estimates of the PAMELA and AMS sensitivities
  to exotic antiprotons for 3 years missions, respectively.}
\label{fig:02}
\end{figure*}

As shown in Ref. \cite{ind}, among the various searches for indirect signals due to annihilation of
light WIMPs, the cosmic rays antiprotons provide the most significant constraints. For this reason
we now examine how this kind of limits applies to the light neutralinos singled out by the DAMA/NaI
annual modulation regions.

So-called secondary antiprotons are produced in the Galaxy via interaction of proton and helium
cosmic rays with the interstellar hydrogen and helium nuclei. A thorough calculation of the
secondary antiproton spectrum has been performed in Ref.\cite{don}, where the antiprotons
generated by spallation processes are propagated using a two--zone diffusion model described in terms
of five parameters. Two of these parameters, $K_0$ and $\delta$, enter the expression of the
diffusion coefficient:
\begin{equation}
K = K_0 \beta R^{\beta},
\label{iso}
\end{equation}

\noindent
where $R$ is the particle rigidity. The other three parameters are the Alf\'en
velocity $V_A$, the velocity of the convective wind $V_c$, and  the thickness
$L$ of the two large diffusion layers which sandwich the thin galactic disk. In
Ref. \cite{don} it is shown that the experimental antiproton spectrum is fitted
quite well by the secondary component from cosmic--rays spallation
(with a ${\chi}^2$ = 33.6 with 32 data points),
calculated with the set of the diffusion parameters which is derived from the
analysis of the boron--to--carbon ratio (B/C) component of cosmic--rays. The
values of this set of best--fit parameters (denoted as median), together with their 4$\sigma$
uncertainty intervals, are given in Table I. The theoretical uncertainty on the
diffusion parameters reflects into a (10 - 20)\% uncertainty on the calculated spectrum
of secondary antiprotons.


\begin{table}[t]
\begin{center}
{\begin{tabular}{@{}|c|c|c|c|c|c|@{}}
\hline
~~{\rm case}~~ &  ~~~~$\delta$~~~~  & $K_0$                 & $L$   & $V_{c}$       &
$V_{A}$
\\  & & ~[${\rm kpc^{2}/Myr}$]~ & ~[kpc]~ & ~[km s$^{-1}$]~ & ~[km s$^{-1}$]~ \\
\hline
\hline
{\rm max} &  0.46  & 0.0765 & 15 & 5    & 117.6  \\
{\rm med} &  0.70  & 0.0112 & 4  & 12   &  52.9  \\
{\rm min} &  0.85  & 0.0016 & 1  & 13.5 &  22.4  \\
\hline
\end{tabular}}
\caption{ Astrophysical parameters of the two--zone diffusion model for galactic
cosmic--rays propagation, compatible with B/C analysis \cite{don} and yielding
the maximal, median and minimal primary antiproton flux.
\label{table:prop}}
\end{center}
\end{table}


Primary antiproton fluxes can be generated by annihilation of neutralino pairs. We have evaluated
these fluxes for the supersymmetric configurations selected by the annual modulation regions,
{\it i.e.} the light neutralino populations which stay inside the annual modulation regions
displayed in the insets of Fig. \ref{fig:01}. These correspond to the cases:
A) $v_0 = 170$ km sec$^{-1}$, $\rho_0 = \rho_0^{\rm min} = 0.20$ GeV cm$^{-3}$;
B) $v_0 = 170$ km sec$^{-1}$, $\rho_0 = \rho_0^{\rm max} = 0.42$ GeV cm$^{-3}$;
C) $v_0 = 220$ km sec$^{-1}$, $\rho_0 = \rho_0^{\rm min} = 0.34$ GeV cm$^{-3}$.

The antiproton fluxes originated in the dark halo by the neutralino
pair-annihilation processes have then been propagated in the diffusive halo
using the three sets of diffusion parameters (minimal, median and maximal) given
in Table I. The procedure for the evaluation of these fluxes is the
one illustrated in Refs. \cite{pbar,ind,ind1}. As it was shown in Ref.
\cite{pbar}, the uncertainty in the diffusion/propagation parameters, contrary
to the case of the secondary antiprotons, induces a large uncertainty on the
primary flux.

To show quantitatively how the experimental data can constrain our
supersymmetric configurations, in Fig. \ref{fig:02} we display the
(top--of--the--atmosphere) antiproton flux evaluated at a specific value of the
antiproton kinetic energy, $T_{\bar{p}} =$ 0.23 GeV, for the three populations A,
B, and C defined above. The shaded (yellow) region denotes the amount of primary
antiprotons which can be accommodated at $T_{\bar{p}} =$ 0.23 GeV without
entering in conflict with the BESS experimental data \cite{bess} and secondary
antiprotons evaluations \cite{don}. The dashed horizontal line denotes our
estimated sensitivity of the PAMELA detector \cite{pam} to exotic antiprotons
after a 3 years running: it corresponds to the admissible excess within the
statistical experimental uncertainty if the measured antiproton flux consists
only in the background (secondary) component. The estimate has been performed by
using the background calculation of Ref. \cite{don}, and refers to a 1--$\sigma$
statistical uncertainty. All the supersymmetric configurations in Fig. \ref{fig:02} above
the dashed line can be potentially identified by PAMELA as a signal over the
secondaries, while those which are below the dashed curve will not contribute
enough to the total flux in order to be disentangled from the background. The
dotted horizontal line represents a similar estimate, but referred to the AMS
detector \cite{ams1} for a 3 years data--taking. Crosses (red) and dots (blue)
denote neutralino configurations selected by the annual modulations regions and
with $0.092 \leq \Omega_{\chi} h^2 \leq 0.124$ and $\Omega_{\chi} h^2 < 0.092$,
respectively. Faint (gray) dots represent configurations which are outside of
the annual modulation regions. Fig. \ref{fig:02} shows that, for values of the diffusion
parameters close to the minimal set, present antiprotons data do not set
constraints.  For values of the diffusion parameters around the median set, we
have that: in case A, most of the neutralino configurations with $\Omega_{\chi}
h^2 < 0.092$ and a few with $0.092 \leq \Omega_{\chi} h^2 \leq 0.124$ remain
unconstrained; in cases B and C only subsets of neutralinos with $\Omega_{\chi}
h^2 < 0.092$ survive. When the diffusion parameters approach the values of the
maximal set, only very few SUSY configurations survive in case A.
From Fig. \ref{fig:02} we see that the possibility of exploring our relevant neutralino configurations by
future measurements of galactic antiprotons (PAMELA and AMS) is quite good.
As was remarked in Sect. II, in case the channelling is not included, 
the fit of the experimental data of annual modulation with light 
neutralinos would  imply values of $\rho_0$ higher than the those 
characterizing the sets A, B and C, previously defined. This property, 
together with the fact that the antiproton flux depends on the square of 
$\rho_0$, might cause tension between the annual modulation data and the 
constraints implied by present measurements of  galactic antiprotons.

\begin{figure}[t] \centering
\vspace{-80pt}
\includegraphics[width=1.15\columnwidth]{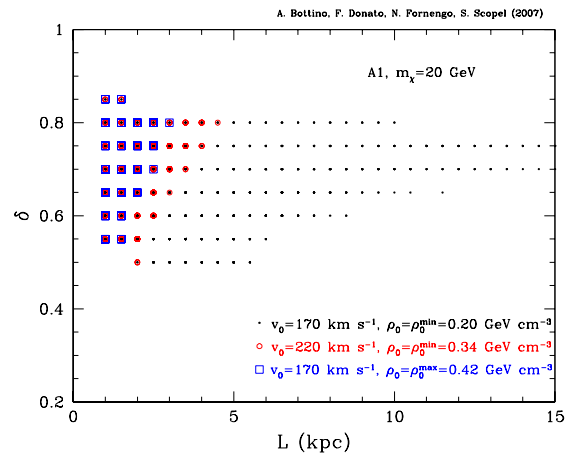}
\vspace{-30pt}
  \caption{Areas of compatibility between the annual modulation regions of Fig. \ref{fig:01} and the
  antiproton data for a neutralino of mass of 20 GeV, plotted in the parameter space defined by the height
  of the diffusive halo $L$ and the rigidity--dependence parameter $\delta$ of the diffusion coefficient
  of Eq. (\ref{iso}). The galactic halo model is a cored-isothermal sphere.
  The dots, (red) squares and (blue) circles refer to:
  $v_0 = 170$ km sec$^{-1}$, $\rho_0 = \rho_0^{\rm min} = 0.20$ GeV cm$^{-3}$,
  $v_0 = 220$ km sec$^{-1}$, $\rho_0 = \rho_0^{\rm min} = 0.34$ GeV cm$^{-3}$,
  and
  $v_0 = 170$ km sec$^{-1}$, $\rho_0 = \rho_0^{\rm max} = 0.42$ GeV cm$^{-3}$, respectively.
  Each set of points shows the region in the $L$--$\delta$ plane which fits at 99.5\% C.L. the
  antiproton data of BESS \cite{bess}.}
\label{fig:fit}
\end{figure}

In the previous analysis, we have considered the antiproton flux evaluated at
a specific value of the antiproton kinetic energy, $T_{\bar{p}} =$ 0.23 GeV.
This analysis can be extended by examining the properties of a global fit to
the full low--energy antiproton spectrum, using the same procedure which was used
in Ref. \cite{ind1}.  We mentioned above that this spectrum is fitted quite well
by the secondary component from cosmic--rays spallation \cite{don}.
As a conservative criterion for constraining our supersymmetric configurations, we can
perform a $\chi^{2}$ analysis on the antiproton data for the supersymmetric configurations
compatible with the annual modulation study. As an illustrative application of this analysis,
for the supersymmetric configurations compatible with each set A, B and C, we calculate
the antiproton flux for all the propagation parameter combinations which keep
the B/C fit within a 4$\sigma$ of uncertainty \cite{bc}.
The primary and secondary fluxes are then required to fit the antiproton data at 99.5\% C.L.
($\chi^2<$60). The results of this calculation are displayed in
Fig. \ref{fig:fit} for $m_{\chi}$ = 20 GeV in the plane of the two diffusion parameters
$L$ and $\delta$. We see that, depending on the specific isothermal--sphere
parameters, we identify different regions of compatibility of the antiproton signal in terms of the
astrophysical parameters which govern the diffusion of galactic cosmic--rays.

\section{Galactic antideuterons}

\begin{figure*}[t] \centering
\vspace{-20pt}
\includegraphics[width=2.0\columnwidth]{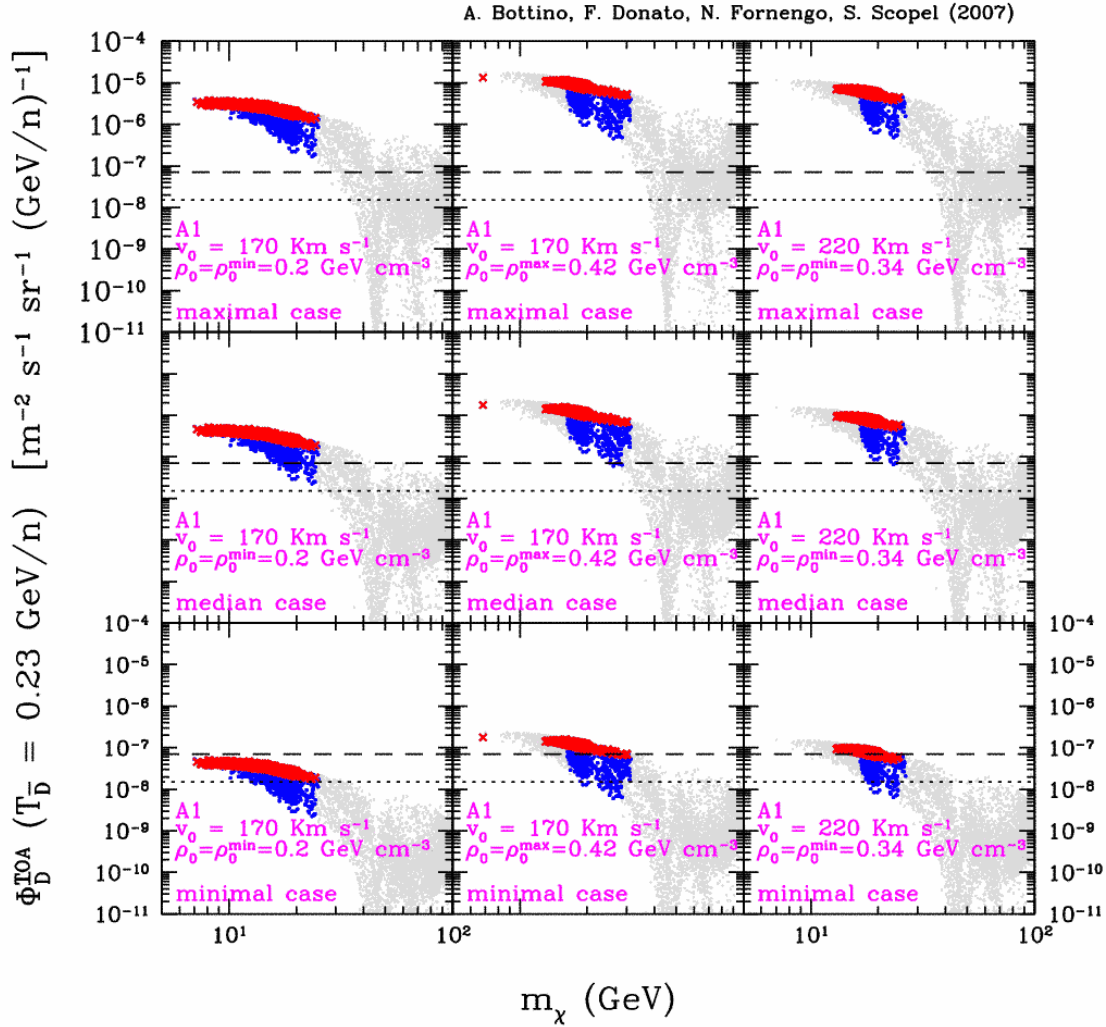}
\vspace{-30pt}
  \caption{Antideuteron flux at $\bar D$ kinetic energy $T_{\bar D}=0.23$ GeV/n,
  as a function of the WIMP mass and for a cored isothermal halo. Notations are
  as in Fig. \ref{fig:02}, except for the horizontal lines, which here refer to estimated
  sensitivities to antideuterons of the GAPS (dashed) and AMS (dotted) detectors.}
\label{fig:03}
\end{figure*}

Formation of antideuterons in cosmic rays proceeds through production
of an antiproton and an antineutron pair by spallation (secondary production) or
by WIMP pair annihilation (primary production) \cite{dbar}. The coalescence process of
antiproton and  antineutron is easier in WIMP annihilation, since their parent particles are
at rest in the galactic frame. Therefore, at low energies the primary
spectrum is much enhanced as compared to the secondary one \cite{dbar}. This feature makes
the search for antideuterons particularly attractive for an  indirect investigation
of WIMPs \cite{dbar,bp}.

The fluxes of the antideuterons produced in the dark halo by the neutralino pair-annihilation
processes have been calculated following the method described in Ref. \cite{dbar} and
propagated in the diffusive halo using the three sets of diffusion
parameters (minimal, median and maximal) given in Table I.

In Fig. \ref{fig:03} we display the (top--of--the--atmosphere) antideuteron flux
evaluated at the
value  $T_{\bar{p}} =$ 0.23 GeV/n for the three population A, B, and C
defined above. The notations for the scatter plot are as in Fig. \ref{fig:02}, that is:
crosses (red) and dots (blue) denote neutralino
configurations selected by the annual modulations regions and with
$0.092 \leq \Omega_{\chi} h^2 \leq 0.124$ and $\Omega_{\chi} h^2 < 0.092$, respectively;
faint (gray) dots represent configurations which are outside of the annual modulation regions.
The horizontal dashed and dotted lines denote estimated sensitivities to antideuterons
of the GAPS \cite{gaps} and AMS \cite{dbar} detectors.

Fig. \ref{fig:03} shows that measurements of galactic antideuterons are perspectively very
promising for investigating our light neutralino populations. Moreover, when
antideuteron data will become available together with the antiproton ones,
correlations in the two data sets will provide strong confidence in a possible
presence of a signal, as can be appreciated by comparing Figs. 2 and 3: for most
of our relevant light--mass neutralinos, a signal should be present both in the
antiproton and antideuteron channel.

\section{Conclusions}

In the present paper we have considered the annual modulation
regions which the DAMA Collaboration has recently determined, by
including also the channelling effect which occurs when an ion
traverses a detector with a crystalline structure, such the detector
of the DAMA/NaI experiment. The inclusion of the channelling effect
implies that the annual modulation region is considerably modified
as compared to the one derived without including channelling. The
extent of the modification depends on the specific model--dependent
procedure employed in the evaluation of the channeling effect.

In the present paper we have considered the phenomenological
consequences for light neutralinos when the annual modulation region
includes the channelling effect as modelled in Ref.\cite{damalast}.
We have proved that these annual modulation data are fitted by light
neutralinos which arise naturally in supersymmetric models where
gaugino mass parameters are not related by the a GUT--scale
unification condition.

The precise range of the neutralino mass which fits the annual modulation data
depends on how the WIMP galactic distribution function is modelled and on a number of
other assumptions, such as those mentioned in Sect. II. As an
example, we have worked out in detail the case of a cored isothermal sphere DF. For this instance,
the neutralino mass stays in the range $m_{\chi} \simeq (7 - 30)$ GeV,
for values of the local rotational velocity, $v_0$, and of the local
dark matter density, $\rho_0$,  in the low-medium
side of their own physical ranges, {\it i.e.} $v_0 \simeq$ (170 - 220) km sec$^{-1}$
and $\rho_0 \simeq (0.2 - 0.4)$ GeV cm$^{-3}$. Similar ranges are found also in
the case of a Navarro--Frenk--White profile or for an isothermal model
with a non-isotropic velocity dispersion.

We have then shown that the populations of light neutralinos selected by the annual
modulation regions are consistent with present data on galactic antiprotons. We have
also derived the intervals of the diffusion parameters which provide this agreement
in correlation with the specific galactic halo model. For instance, for neutralinos
with a mass of 20 GeV and a cored isothermal model with $v_0 = 170$ km s$^{-1}$ we have
$0.55 \lsim \delta \lsim 0.85$ and $L \lsim 3$ kpc when $ \rho_0 = \rho_0^{\rm max} = 0.42$ GeV cm$^{-3}$;
instead when $ \rho_0 = \rho_0^{\rm min} = 0.20$ GeV cm$^{-3}$, $L$ may go up to 15 kpc with a range of
$\delta$ which progressively shrinks to $\delta \sim 0.70 - 0.75$, when $L$ increases.

We have also shown that future measurements of galactic
antiprotons and antideuterons
will offer, together with the
upcoming data from DAMA/LIBRA,  very interesting perspectives for further investigating
the light neutralino
populations selected by the annual modulation data. In case of models with a corotating halo or with
triaxial spatial distributions, not investigated in the present paper,
also heavier neutralinos can be involved.

Finally, a word of caution should be said concerning the fact that the distribution
of WIMPs in the Galaxy could deviate from the models mentioned above, mainly
because of the presence of streams and/or clumpiness. In such instances, the analysis
should be appropriately adapted, along the lines discussed in the present paper.

\acknowledgments
We thank the DAMA Collaboration  for informing us of its work prior to
publication. We are also grateful to Rita Bernabei and
 Pierluigi Belli for useful discussions.
 We  acknowledge Research Grants funded jointly by Ministero dell'Istruzione,
dell'Universit\`a e della Ricerca (MIUR), by Universit\`a di Torino and by
Istituto Nazionale di Fisica Nucleare within the {\sl Astroparticle Physics
Project}.

\medskip
\section{Appendix: The supersymmetric model}
\label{sec:susy}

The supersymmetric scheme we employ in the present paper is the
one described in Ref.  \cite{lowneu}: an effective MSSM scheme
(effMSSM) at the electroweak scale, with the following independent
parameters: $M_2, \mu, \tan\beta, m_A, m_{\tilde q}, m_{\tilde l},
A$ and $R \equiv M_1/M_2$. Notations are as
follows: $\mu$ is the Higgs mixing mass parameter, $\tan\beta$ the
ratio of the two Higgs v.e.v.'s, $m_A$ the mass of the CP-odd
neutral Higgs boson, $m_{\tilde q}$ is a squark soft--mass common
to all squarks, $m_{\tilde l}$ is a slepton soft--mass common to
all sleptons, $A$ is a common dimensionless trilinear parameter
for the third family, $A_{\tilde b} = A_{\tilde t} \equiv A
m_{\tilde q}$ and $A_{\tilde \tau} \equiv A m_{\tilde l}$ (the
trilinear parameters for the other families being set equal to
zero).

Since we are here interested in light neutralinos, we consider
values of $R$ lower than its GUT value: $R_{GUT} \simeq 0.5$. For
definiteness, we take $R$ in the range: 0.005 - 0.5.

In the present paper the numerical analyses are performed by a
scanning of the supersymmetric parameter space, with the following
ranges of the
MSSM parameters: $30 \leq \tan \beta \leq 50$, $100 \, {\rm GeV}
\leq |\mu| \leq 300 \, {\rm GeV}, 100 \, {\rm GeV}
\leq M_2 \leq 1000 \, {\rm GeV},
100 \, {\rm GeV} \leq m_{\tilde q}, m_{\tilde l} \leq 1000 \, {\rm GeV }$,
$90\, {\rm GeV }\leq m_A \leq 150 \, {\rm GeV }$, $-3
\leq A \leq 3$.

The following experimental constraints are imposed: accelerators data on
supersymmetric and Higgs boson searches (CERN $e^+ e^-$ collider LEP2
\cite{LEPb} and Collider Detectors D0 and CDF at Fermilab \cite{cdf});
measurements of the $b \rightarrow s + \gamma$ decay process \cite{bsgamma}:
2.89 $\leq B(b \rightarrow s + \gamma) \cdot 10^{-4} \leq$ 4.21 is employed
here: this interval is larger by 25\% with respect to the experimental
determination  \cite{bsgamma} in order to take into account theoretical
uncertainties in the SUSY contributions  \cite{bsgamma_theorySUSY} to the
branching ratio of the process (for the Standard Model calculation, we employ
the recent NNLO results from Ref.  \cite{bsgamma_theorySM}); the upper bound on
the branching ratio $BR(B_s^{0} \rightarrow \mu^{-} + \mu^{+})$ \cite{bsmumu}: we
take $BR(B_s^{0} \rightarrow \mu^{-} + \mu^{+}) < 1.2 \cdot 10^{-7}$;
measurements of the muon anomalous magnetic moment $a_\mu \equiv (g_{\mu} -
2)/2$: for the deviation $\Delta a_{\mu}$ of the  experimental world average
from the theoretical evaluation within the Standard Model we use here the range
$-98 \leq \Delta a_{\mu} \cdot 10^{11} \leq 565 $, derived from the latest
experimental  \cite{bennet} and theoretical \cite{bijnens} data.


\begin{thebibliography}{99}
\bibitem{nota1} In deriving the regions of Fig. 1, cases A, B, C defined in Sect. 7.2 of
the first paper of Ref. \cite{dama} have been included.
\bibitem{nota2} In deriving the regions of Fig. 2, only case A defined in Sect. 7.2 of
the first paper of Ref. \cite{dama} has been included.
\end{thebibliography}
\end{document}